# Memristive Nanowire Network for Audio Classification: Pre-Processing-Free Neuromorphic Computing with Reduced Latency

**Akshaya Rajesh[1], Pavithra Ananthasubramanian[1], Nagarajan Raghavan[1], Ankush Kumar[1,2]***

[1] nano-Macro Reliability Laboratory (nMRL), Engineering and Product Development Pillar, Singapore University of Technology and Design, 8, Somapah Road, 487372, Singapore.

[2] Centre for Nanotechnology, Indian Institute of Technology Roorkee, Roorkee, Uttrakhand, 247667, India

*Corresponding author email: ankush.kumar@nt.iitr.ac.in

**Abstract**

Efficient audio feature extraction is critical for low-latency, resource-constrained speech recognition. Conventional preprocessing techniques, such as Mel Spectrogram, Perceptual Linear Prediction (PLP), and Learnable Spectrogram, achieve high classification accuracy but require large feature sets and significant computation. The low-latency and power efficiency benefits of neuromorphic computing offer a strong potential for audio classification. Here, we introduce memristive nanowire networks as a neuromorphic hardware preprocessing layer for spoken-digit classification, a capability not previously demonstrated. Nanowire networks extract compact, informative features directly from raw audio, achieving a favorable trade-off between accuracy, dimensionality reduction from the original audio size (data compression) , and training time efficiency. Compared with state-of-the-art software techniques, nanowire features reach 98.95% accuracy with ~66 times data compression (XGBoost) and 97.9% accuracy with 255 times compression (Random Forest) in sub-second training latency. Across multiple classifiers nanowire features consistently achieve >90% accuracy with ≥62.5 times compression, outperforming features extracted by conventional state-of-the-art techniques such as MFCC in efficiency without loss of performance. Moreover, nanowire features achieve 96.5% accuracy classifying multispeaker audios, outperforming all state-of-the-art feature accuracies while achieving the highest data compression and lowest training time. Nanowire network preprocessing also enhances linear separability of audio data, improving simple classifier performance and generalizing across speakers. These results demonstrate that memristive nanowire networks provide a novel, low-latency, and data-efficient feature extraction approach, enabling high-performance neuromorphic audio classification.

## 1. Introduction

The rapid proliferation of edge devices and multimodal data streams has amplified the demand for energy-efficient and low-latency artificial intelligence. In domains such as healthcare, robotics, and smart sensing, massive data transfer to cloud infrastructures introduces prohibitive energy costs and latency constraints[1-4]. Although deep neural networks (DNNs) have transformed machine learning with remarkable accuracy across perception and reasoning tasks[5], their performance comes at the expense of enormous computational overhead and energy consumption that scales unfavourably with model size and training data volume[6-8]. In contrast, the human brain executes complex sensory processing, such as auditory recognition, with approximately 10–20 W of power[9, 10], motivating intense interest in neuromorphic systems that replicate this efficiency through in-memory, massively parallel computation.

Neuromorphic computing leverages physics-based systems that emulate neural behaviour to overcome the von Neumann bottleneck, offering co-located processing and storage at the device level[3, 11-14]. Among the diverse approaches proposed, memristive technologies have shown particular promise for realizing energy-efficient and adaptive architectures. Memristors implemented in crossbar arrays or multi-terminal geometries, using materials such as resistive switching memory[15], non-volatile random access memory [3], NAND flash memory architecture [16], and STT-RAM crossbar arrays [17], have been successfully applied to image classification[18, 19], pattern recognition[20], and spiking computation[12, 21]. Various neuromorphic devices based on nano-scale materials have also been proposed with two and three-terminal architectures [21] utilizing a range of materials including organic materials [22], perovskites [23], quantum dots [24] etc. However, most existing architectures require explicit device fabrication into predefined topologies, limiting scalability and adaptive connectivity.

Self-assembled memristive nanowire networks (NWNs) offer an alternative paradigm. These networks form through bottom-up assembly of nanowires whose junctions act as tunable, memristive synapses, enabling spontaneous connectivity, recurrent dynamics, and local plasticity[25-31] Their emergent organization gives rise to a natural substrate for *in-materio* computation, where temporal signals are processed and stored within the physical dynamics of the material itself[25, 29, 30]. Memristive nanowire networks (NWNs) show significant similarities to the signal storage and processing expressed by neurons in the mammalian brain with synapse-like connections and neuron-like junctions. [25-31]. Various reports have studied memristive materials in cross-bar array configurations [32] for data classification in detail and have used them for several tasks such as image classification [18, 19], facial recognition in images [33], pattern recognition [20], natural language processing [34], game playing [35], and MNIST spoken digit classification [36-41]. Recent studies have harnessed NWNs, spiking neural networks and atomic switch networks (ASNs) for spoken-digit classification[25, 37, 39, 40], demonstrating computational capabilities in-materio to transform temporal data.

Nanowire network-based approaches are especially impactful for speech or audio classification due to their memristive real-time computation of temporal data. Treating data through spiking neural networks before using linear classifiers[37, 40] and atomic switch networks (ASN) incorporating silver iodide (AgI) junctions[25] are proven neuromorphic approaches that achieve 90% classification accuracy.

However, there is still a large reliance on computational heavy software preprocessing, either in the form of spiking neural networks or more well-known audio processing techniques such as Mel-frequency cepstral coefficients (MFCCs), spectrograms, or wavelet transforms. While spiking neuromorphic approaches have achieved competitive accuracies on spoken-digit datasets [36-41], none have yet demonstrated direct hardware-level feature extraction from raw audio signals. Bridging this gap requires a system capable of transforming raw temporal waveforms into compact, discriminative feature representations entirely through its internal dynamics, without any software intervention.

Here, we propose a memristive nanowire-network (NWN)–based neuromorphic preprocessing framework that performs *in-materio* feature extraction directly from raw audio waveforms. Unlike prior approaches dependent on digital preprocessing, our system leverages the intrinsic spatiotemporal transformations of the nanowire network, arising from its memristive junction dynamics and recurrent architecture, to project raw temporal data into a high-dimensional conductance space. This process naturally compresses, denoises, and linearizes the input signal while maintaining the discriminative structure required for downstream classification. We further integrate a subsampling strategy to quantify and optimize feature compactness, enabling quantitative evaluation of the trade-offs between feature dimensionality, classifier accuracy, and computational latency. The study compares the classification accuracy of using features extracted by the nanowire network against those extracted by state-of-the-art audio preprocessors for spoken-digit classification, considering single-speaker and multi-speaker classification use cases to assess the generalizability of nanowire-extracted features.

By comparing the impact features extracted by nanowire networks against state-of-the-art software audio feature extractors, this study presents a novel hardware feature extraction technique and compares its effectiveness across single-speaker and multispeaker datasets. It delves into the impact reservoir computing transformations on temporal data, and systematically investigates the impact of nanowire-network parameters, on spatio-temporal transformations of audio signals.

These contributions present a pathway toward neuromorphic systems capable of real-time, energy-efficient auditory perception at the edge. By demonstrating that memristive nanowire networks can autonomously extract and compress meaningful representations from raw temporal signals, this work positions NWNs as a scalable foundation for next-generation cognitive hardware.

## 2. Method

In this study, memristive nanowire networks are explored for an audio signal classification task, leveraging the unique spatiotemporal transformation they offer to enhance the performance of software classifiers. Figure 1 is a schematic of the processes involved in the audio classification task using a nanowire network model discussed in this study.

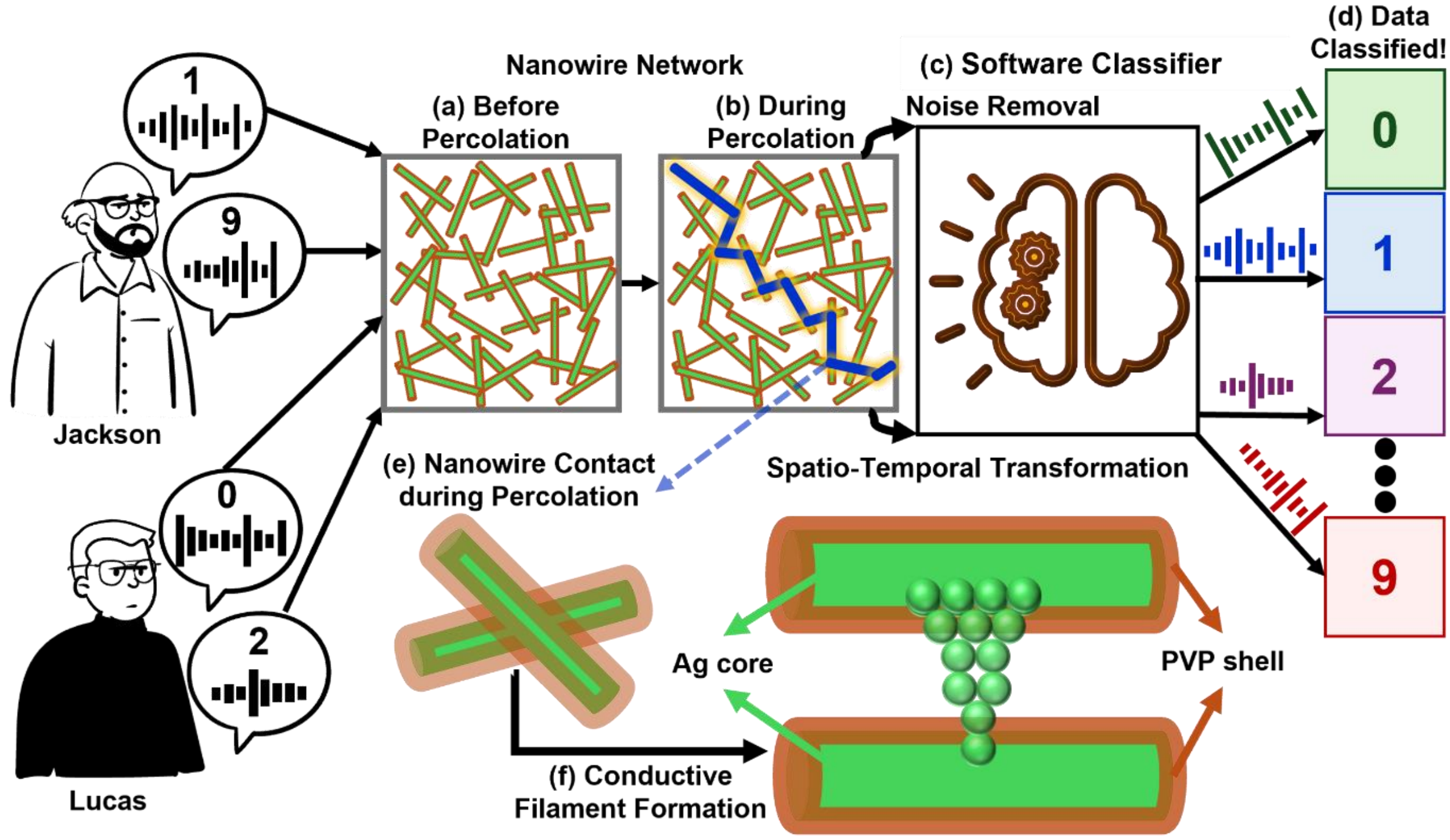

*Figure 1: Graphical abstract of study conducted with the nanowire networks. Data from multiple speakers were used from the audio dataset for 10-class classification, by passing the data through the nanowire network. The audio signals formed percolation paths through the network, and the conductance measurement was taken at the drain point and passed as input into the software classifier for the final classification. Image is not to scale and is only for representative purposes.*

*2.1 Nanowire Network Simulation*

The data processing using a nanowire network was conducted using a custom-built simulation in Python. This Python program emulates a nanowire network's signal processing capabilities, comprising 1500 nanowires with mean lengths of 40nm (std. = 14nm), self-assembled on a substrate.

This network is realized in hardware by randomly drop-casting memristive nanowires on an insulating substrate. Nanoscale cross-point junctions at the intersections of the nanowires form memristive cells where the internal state of resistance depends on the history of electrical stimuli. [42]The nanowires form a highly interconnected self-realized network, with junctions between wires exhibiting variable conductance and memristive properties.[43]. When the amplitude of an audio file is passed in as voltage signal inputs, a conductance bridge can form across the individual nanowire junctions, which allows the signal to percolate between conducting nanowires [27, 29, 42, 43]. Based on the conductance of each junction in the network, current applied at a node in the network (source) forms a percolation path to other nodes in the network. In nanowire networks, the percolation path forms through collective switching dynamics where junctions near electrodes with voltages exceeding the threshold ($|V_{jn}| > V_{set}$) begin growing filaments, triggering voltage redistribution to neighbouring junctions through Kirchhoff's laws, causing the network to self-organize and funnel current down the shortest pathway that forms first in a competitive growth process from electrodes toward the network centre. [26]. The output voltage is measured at one of these nodes (chosen ground node) and the path the current takes to travel from the source to ground is known as the percolation path. The current distribution and percolation paths is modelled with Kirchoff's current laws and dependant on the conductance of each junction in the network.

Due to their memristive property, the conductance of each junction increases or decreases based on the potentiation or depression occurring. The equivalent resistance between the source and the ground depends on the resistance states of all the junctions in the network. The path taken by the currents through the network are governed by Kirchoff Law equations. This output conductance is monitored over time intervals for each audio file, creating a conductance time series that represents the audio signal for classification tasks. The simulation allows for the configuration of several input hyperparameters, governing two main properties of the network: the memristive junctions and the recurrent connections.

**Memristive Junctions**

The conductance of these junctions is controlled by two key hyperparameters: potentiation and depression rates. These time constants dictate the formation and dissolution of the conductance bridges in response to an input signal. The relation between the junction conductance and the rates are as expressed by the memory state equation [42] as follows.

$$\frac{dg}{dt} = K_p(V)(1-g) - K_d(V)g \quad (1)$$

$$K_{(p,D)} = k_{(p,D)} exp(\eta_{(p,D)}, V) \quad (2)$$

Here, $k_{(p,D)}$ represents the potentiation and depression rates, which are set in the simulation. The junction conductance is regulated by these hyperparameters, and the voltage signals applied during previous timesteps. During processing, a new network is instantiated for each audio file to ensure that the conductance at any time step is influenced only by previous inputs from the same signal.

**Recurrent Connections**

The network-level properties, such as size, density, and wire length, are fixed variables in the simulation. These parameters guide the Monte Carlo-based self-assembly of the network. Nanowire networks possess an inherent recurrent network structure where multiple nanowires form interconnected junctions, creating feedback loops that enable the network to exhibit brain-like collective dynamics and memory properties. The feedback loops within the network influence transformations to the percolating signal, while the output signal can be measured instantaneously at any junction point. When an audio signal is passed through, it activates percolation pathways, or current-carrying backbones, modelled using Kirchhoff's voltage laws. Multiple percolation pathways may be activated due to the recurrent nature of the connections. Therefore, the source and ground nodes are positioned at the extreme corners of the network (~7 times the length of a single nanowire) to observe these multi-scale pathways effectively. The resultant conductance time series from the nanowire network simulation was subsequently used as input for classification tasks, linking the network's signal processing capabilities to the overall model performance (Figure 1).

*2.2 Nanowire Network Parameter Optimisation*

Preliminary parameter selection for the nanowire network was conducted by processing an audio file (George Digit 0 Trial 0) through the network while systematically varying key network parameters. The output conductance was measured at a defined ground node and visualised to assess the network's performance in noise suppression and signal preservation. The three parameters optimized in this study were the potentiation rate ($K_p$), depression rate ($K_d$), and voltage amplitude ($V_p$). The potentiation and depression rate define the rate of formation and dissolution of the conductance bridge at the nanowire network node junction, influencing the conductance spike at a particular time step signal. These rates determine the conductance values based on the amplitude of signal history. For example, if the depression rate is low and potentiation rate is high, then it is easier for an upcoming signal to cause a spike since the node junction is more sensitive to potentiation and the spike decay of the previous signal is still not complete. Similarly, voltage amplitude also influences the signal spike strength. By tuning these values, we aim to bring about a smoothing effect over the original audio file, for which setting appropriate $K_p$, $K_d$, and $V_p$ values is critical.

A preliminary range for each parameter was determined based on prior research and initial exploratory tests. $K_p$ was tested in the range of [0.0001-0.5], $K_d$ in the range of [0.3-0.5], and $V_p$ in the range of [0.5-5]. The optimization process began with $K_p$, keeping $K_d$ and $V_p$ constant. The same audio file was input into the network for each iteration, with $K_p$ varied systematically. The output conductance at the ground node was measured and plotted. Values of $K_p$ above 0.001 led to signal saturation, showing no discernible trends, thus narrowing the relevant $K_p$ values to 0.001 or 0.0001. Subsequently, these two $K_p$ values were tested with varying $K_d$ values while maintaining a constant $V_p$. The same procedure was followed, with the output conductance plotted for each combination. The optimal $K_p$ and $K_d$ values were selected based on their ability to minimize noise without compromising the signal's integrity. Finally, with $K_p$ and $K_d$ set to their optimal values, $V_p$ was varied within the chosen range. The audio file was processed for each $V_p$ value, with the output conductance recorded and analyzed. The final parameter set was selected based on its effectiveness in noise reduction while preserving the signal. The preliminary parameter selection process is crucial in ensuring the network's ability to enhance noise suppression, which directly impacts the classifiers accuracy and overall robustness of the system. The values of $K_p$= 0.001, $K_d$ = 0.5 and $V_p$ = 1 are selected as final parameters and kept constant for every audio file run. The nanowire network is expected to perform two key tasks: (i) noise suppression, and (ii) spatiotemporal transformation. Memristive nanowire networks are known for suppressing high-frequency signals leading to noise reduction. Furthermore, due to their recurrent connections, they non-linearly project the output to other neurons in the network, transforming the temporal signal into a spatial signal. Multiple parameters affect the performance of the network in these 2 tasks, such as potentiation rate, depression rate, nanowire density, voltage amplitude, etc.

We conducted a systematic investigation of the potentiation rate ($K_p$) in the range [0.001-0.5], depression rate ($K_d$) in the range [0.3-0.5], and voltage amplitude ($V_p$) in the range [0.5-5]. Figure 2 details the results of this analysis. As shown in Figure 2 (a), values of $K_p$ below 0.01 result in under-potentiation causing a low-dampened signal output. Conversely, values above $K_p$ = 0.01 over-potentiate, causing the signal to saturate. This occurs because the potentiation rate controls the conductance of the nanowire node junctions when signals are passed through, with higher $K_p$ increasing the node's potential to spike. Based on these observations, we selected 0.01 as the optimal $K_p$ value.

Similarly, we analyze $K_d$, the depression rate that controls the rate of decay in the spike. By adjusting the potentiation and depression rates, the noise suppression of the nanowire network can be tuned. This smoothening of the original signal is analogous to moving average filters, a common smoothening technique used in signal processing to enhance underlying important features of the audio signal. An analysis is performed for $K_d$ and $V_p$ (Figure 2 (b), Figure 2(c)), leading to the final chosen values as shown in Figure 2(d).

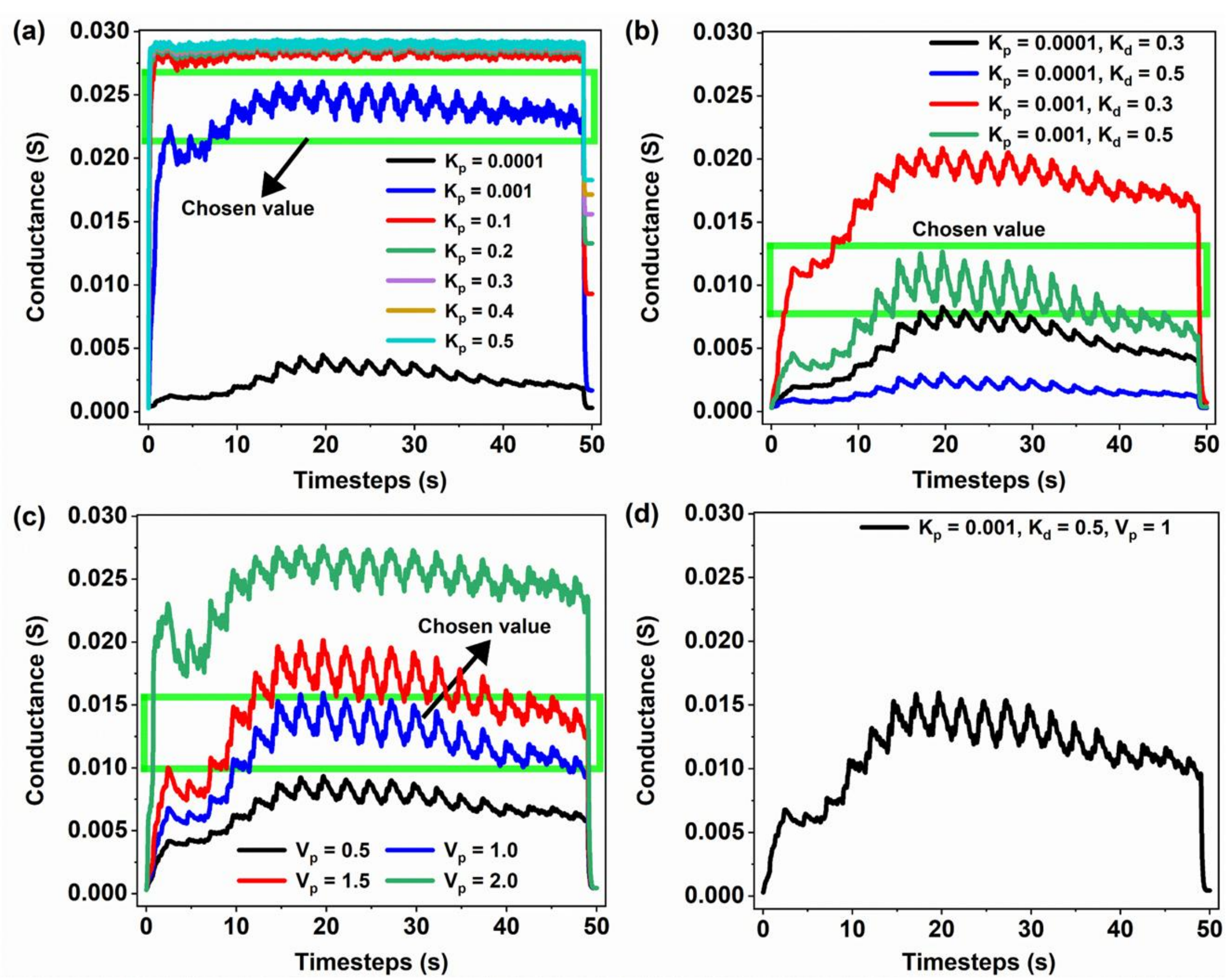

*Figure 2: Optimization of Conductance Parameters for Memristive Nanowire Networks(a) Optimization of $K_p$: The conductance (S) is plotted against timesteps for various values of potentiation rate $K_p$, with the chosen value highlighted.(b) Optimization of $K_d$: Conductance behaviour for different depression rates $K_d$ with the selected optimal value indicated.(c) Optimization of $V_p$: Variation in conductance with different voltage amplitudes $V_p$ at the selected $K_p$ and $K_d$ values, showing the chosen optimal $V_p$.(d) Final Conductance Behaviour: The conductance plot over time using the optimized parameters ($K_p$=0.001, $K_d$=0.5, $V_p$=1) selected from the previous optimizations.*

*2.3 Subsampling Technique*

The nanowire network outputs a 1024-length feature array for each audio file it processes. However, not all 1024 of these features are required to achieve peak classification accuracy, hence the number of features from the network can be reduced. This selection is performed using subsampling, a process in which features at regular intervals are selected while remaining features are discarded. This approach allowed us to cull the input by selectively identifying the data points that provide the most relevant information to classification, thus improving the accuracy and reducing the mean training time of our models (SI.1)

First, we determine the size of the feature subset, which is defined as the number of features retained from the original 1024-timestep array outputted by the nanowire network for an audio clip. Given the temporal nature of nanowire network processing, evenly spaced points were selected in powers of 2, from $2^0$ to $2^{10}$, reflecting the standard 1024-length output array of the nanowire network. Each of the machine learning models were trained on subsampled nanowire data, with the number of features of the highest accuracy configuration used to compute the data compression by nanowires from the original 4087 timestep mean audio sample length.

The testing data is subsampled at the same size as the training data and passed through the classifier to measure the accuracy of classification. The accuracy results are collected tabularly and visualized

*2.4 Classification Tasks*

We evaluated nanowire networks as neuromorphic feature extractors on multiple spoken-digit classification tasks, comparing their performance with three state-of-the-art software preprocessing methods: Mel Spectrogram, Perceptual Linear Prediction (PLP), and Learnable Spectrogram. Features obtained from each method were passed to four nonlinear classifiers, Random Forest (RF), XGBoost, single-layer neural network (1-NN), and two-layer neural network (2-NN). Nanowire-processed outputs were subsampled to different dimensionalities (between $2^0$ to $2^{10}$) before classification. The software classifiers are each optimized for each pre-processed dataset to maximize accuracy. Performance metrics included classification accuracy, data compression ratio (DCR), and training time. Data compression ratio is calculated using the following formula, where 4087 is the mean length of the original audio sample:

$$DCR = \frac{Number\ of\ feature\ passed\ to\ software\ classifier}{4087}$$

To assess the impact of nanowire preprocessing on linear separability, nanowire features were also evaluated with three linear classifiers, support vector machine (SVM), linear discriminant analysis (LDA), and logistic regression (LR) (SI2, SI3).

**Dataset**

Experiments were conducted on the Free Spoken Digit Dataset (FSDD), an open-source corpus of 3,000 recordings of digits 0–9 (50 repetitions per digit, six speakers, 8 kHz sampling rate). Due to computational constraints in nanowire simulations, a subset was used comprising 1,200 recordings: three speakers (George, Lucas, Jackson), 10 digits, and 40 trials per digit. For speaker generalization, Jackson was used for training and Lucas/George for testing.

**Experimental Tasks**

We designed 2 complementary classification tasks to evaluate the effectiveness of nanowire-based feature extraction:

1. Individual-Speaker Nonlinear Classification: For each speaker in the dataset, we performed 10-class digit classification using nonlinear classifiers (Random Forest, XGBoost, 1-layer NN, 2-layer NN) on nanowire-processed features and on features extracted via state-of-the-art software preprocessing methods (Mel Spectrogram, PLP, Learnable Spectrogram). This task evaluates how nanowire preprocessing affects overall classification accuracy at the single-speaker level.
2. Speaker-Generalization Test: To test the speaker-agnostic properties of nanowire features, we performed 10-class classification on a multi-speaker dataset using nonlinear classifiers. Models were trained on one speaker (Jackson) and tested on others (Lucas and George). Classification accuracy and data compression ratio were compared with software-preprocessed features to evaluate the robustness of nanowire-based representations across speakers.

Additionally, experiments testing the effect of nanowire preprocessing on the linear separability of the data were run, and results can be found for single and multi-speaker in SI.2 , SI.3.All classifiers were implemented in Python (scikit-learn) using a 90/10 train–test split and 5-class cross-validation.

### *2.5. Task Difficulty Analysis*

To illustrate and quantify the separability between classes, Euclidean distance is implemented. The Euclidean distance calculates the distance between the two audio samples by taking the straight-line distance between each corresponding point in the time series. The Euclidean distances between two audio files is defined as in Equation (3). Euclidean distance between Audio 1 ($t_{11}$, $t_{12}$..t1n) and Audio 2 ($t_{21}$, $t_{22}$…$t_{2n}$) is represented in equation 3.

$$\text{Euclidean distance} = \sqrt{\Sigma_{i=1}^{n}(t_{1i} - t_{2i})^2} \qquad (3)$$

where t is the timestep and n is the total number of timesteps in the length standardized audio clips.

Audios that have similar acoustic features (Fig 3(a)) such as pitch, duration, and spectral properties tend to be closer together in the Euclidean distance dimension. The phenomes or spoken digits with similar sound characteristics can result in similar feature vectors in the Euclidean dimension as compared to audios of different classes (Fig 3(b)). Higher Euclidean distance between interclass data points and lower distance between intraclass data points signifies higher linear separability and clearer decision boundaries, which linear models can more effectively learn. A lower standard deviation in intraclass Euclidean distance, signifying more consistency between similar audio samples, is also expected.

Figure 3 shows the results of performing a pairwise calculation of the distance between all audio files, taking the mean of the intra and interclass audio files. It was observed that the Euclidean distances for both interclass and intraclass falls between 2 and 3.6 with significant overlap between the intra and interclass values. For example, the intraclass distance for digit 0 is equal to the interclass distance between digits 0 and 6 (Figure 3(c)). Some classes, such as digits 0 and 6, have lower mean intraclass distances than mean interclass distances (Figure 3(d)). The proximity of interclass means to intraclass indicates that even across classes, similar phenomes make distinguishing between them challenging.

Additionally, the standard deviation of the intraclass Euclidean distance is higher than that of the interclass Euclidean distance as shown in Figure 3(d). This illustrates greater variability between phenomes of the same class, adding to the difficulty of the task. The lower intra class variability suggests that classes are, on average, separable however there is significant variability in the raw data that complicates the classification, possibly due to noise. The high overlap in intra and intra class ranges, coupled with high intraclass variability, indicates that interclass audios are not easily distinguishable from intraclass samples, making it challenging for classifiers to establish clean decision boundaries.

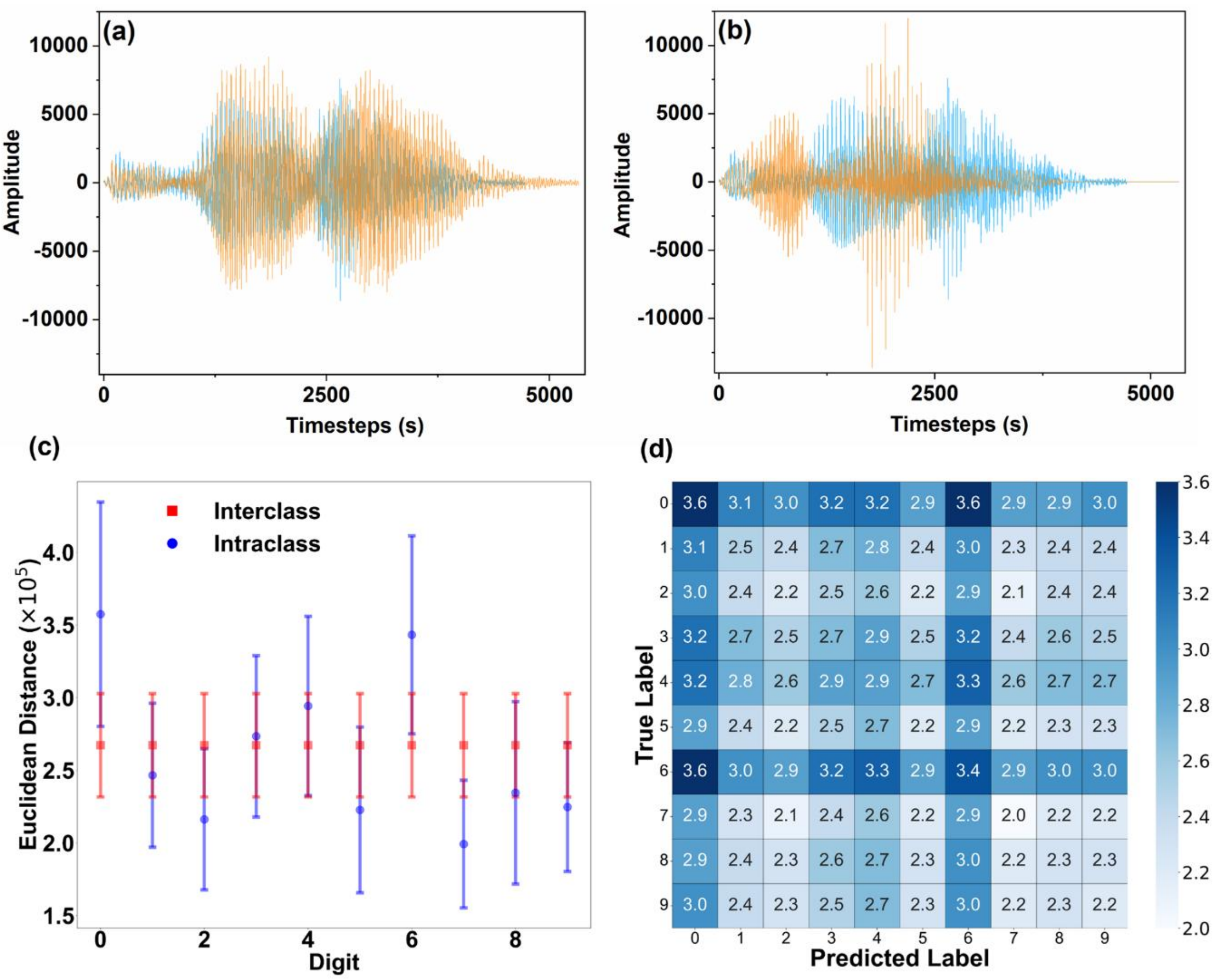

*Figure 3: Analysis of intraclass and interclass audio signal comparisons and their corresponding Euclidean distances.(a) Intraclass Audio Comparison: Superimposed waveforms of two audio samples from the same digit class, illustrating consistent amplitude patterns and waveform similarities.(b) Interclass Audio Comparison: Superimposed waveforms of two audio samples from different digit classes, demonstrating distinct differences in amplitude and waveform structure.(c) Euclidean Distance Analysis: Comparison of average Euclidean distances for intraclass (blue) and interclass (red) audio samples across different digits. Error bars denote standard deviations, indicating variability in the Euclidean distances. (d) Interdigit Euclidean Distance Matrix: Heatmap representing the average Euclidean distances between audio samples from different digit classes demonstrates higher distance values between digit classes (interclass).*

## 3. Results

### *3.1. Classification accuracy v data compression v training efficiency*

Nanowire network preprocessing demonstrates superior feature extraction capabilities, achieving 98.95% classification accuracy with 65.9x data compression (64 features from mean-length of original sample waveform of 4087) using XGBoost, and 97.9% accuracy with 255.4 times compression (16 features) using Random Forest (Figure 4a). These results represent a fundamental advance in neuromorphic audio processing, as nanowire networks extract features that are simultaneously more compact and nearly as informative as conventional software techniques.

Direct comparison with state-of-the-art preprocessing techniques demonstrates nanowire networks' superior compression capabilities. While Mel Spectrogram achieved 98% accuracy with 132 features, PLP reached 99% with 256 features, and Learnable Spectrogram attained 100% with 512 features, while nanowire networks reach comparable or superior accuracy with just 16-64 features. This compression directly translates to computational efficiency, as the nanowire-random Forest pipeline achieved sub-second training time (0.28s) compared to 0.71s for the fastest software alternative (Learnable Spectrogram-Random Forest), while maintaining >97% accuracy (Figure 4b).

The robustness of nanowire features was validated across multiple classifier architectures. Single-layer neural networks achieved 92.83% accuracy with 128 nanowire features, while two-layer networks reached 93.51% with just 16 features, demonstrating that the extracted features remain discriminative even with simple model architectures (Table 1). This classifier-agnostic performance suggests that nanowire preprocessing captures fundamental acoustic patterns rather than classifier-specific artifacts.

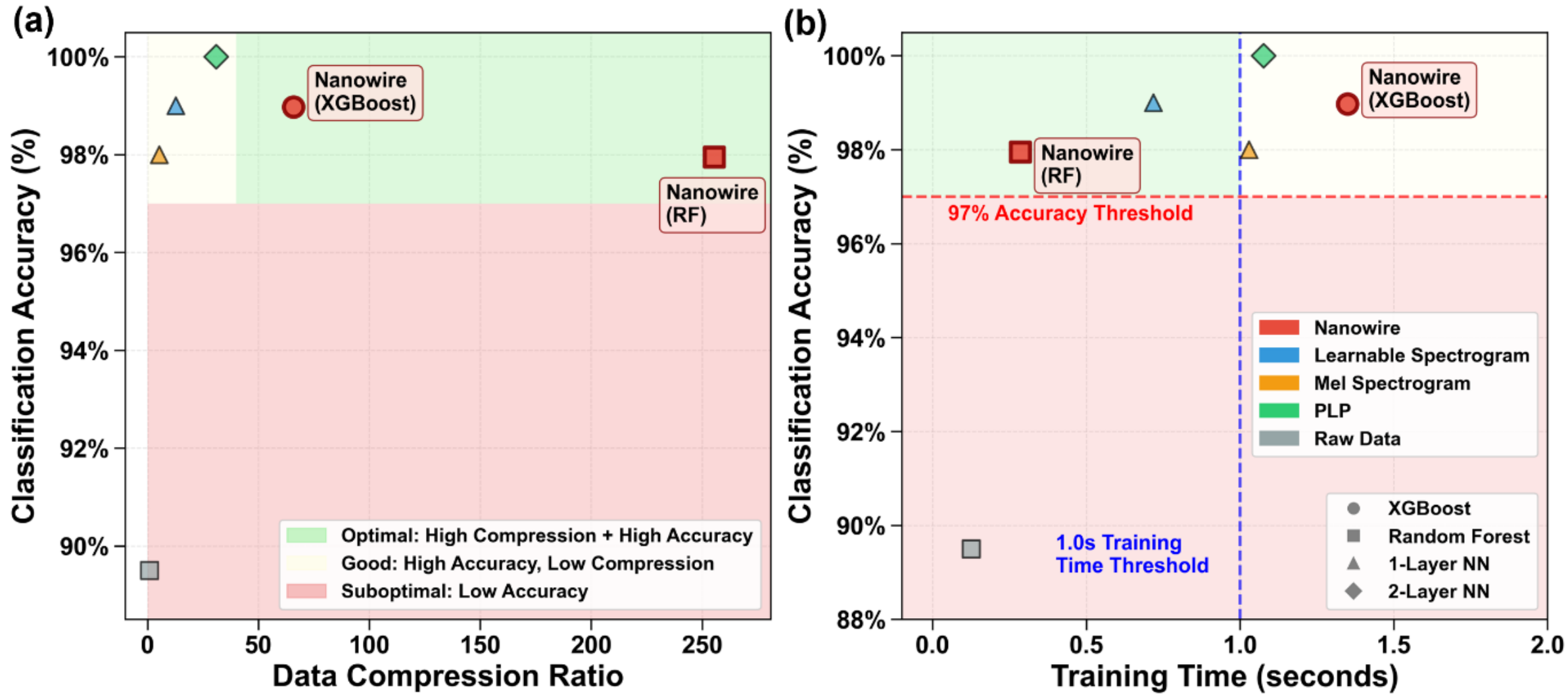

*Figure 4. Combined analysis of classification accuracy, data compression ratio, and training efficiency across feature representations and classifiers. (a): classification performance versus data compression ratio. Data compression ratio is the ratio of features passed to the classifier over the mean-length of original audio samples. Nanowire preprocessing attains up to 255 times compression while maintaining >97%. accuracy (b):accuracy versus training time reveals that nanowire representations enable sub-second training (<1 s) while maintaining accuracy above the 97% threshold. Nanowire–Random Forest achieves the fastest training time (0.28 s) at high compression and accuracy, outperforming raw and spectrogram-based pipelines in both speed and data efficiency.*

| Classifier | Max Accuracy | Max Accuracy: Subset Size | Compression Factor |
|---|---|---|---|
| XGBoost | 98.97% | 64 | 65.9x |
| Random Forest | 97.95% | 16 | 255.4x |
| 1-Layer NN | 92.83% | 128 | 31.9x |
| 2-Layer NN | 93.51% | 16 | 255.4x |

Table 1. Maximum classification accuracy and corresponding feature-set sizes for nanowire-processed data across classifiers. Nanowire features achieve >90% accuracy across all models with compression factors from 31.9 times to ~255 times, outperforming other preprocessing pipelines in efficiency without loss of accuracy.

### *3.2. Multispeaker Classification Accuracy v Data Compression*

To evaluate speaker generalization, we compared nanowire preprocessing against state-of-the-art software techniques on multi-speaker spoken-digit classification using nonlinear classifiers. As shown in Figure 5, nanowire features achieve the highest accuracy (96.5%) while simultaneously offering the greatest data compression (125x) and the lowest training time (1.92s). By contrast, only Learnable Spectrogram surpasses 90% accuracy (95.6%), while Mel Spectrogram and PLP remain below 85%, and raw data classification falls to 42%.

These results indicate that nanowire-extracted features are speaker-agnostic, preserving discriminative information across different speakers despite aggressive compression. The combination of high accuracy, strong generalization, and computational efficiency demonstrates the suitability of nanowire preprocessing for edge-deployed voice recognition systems, where models must adapt to unseen users while operating under strict latency and resource constraints.

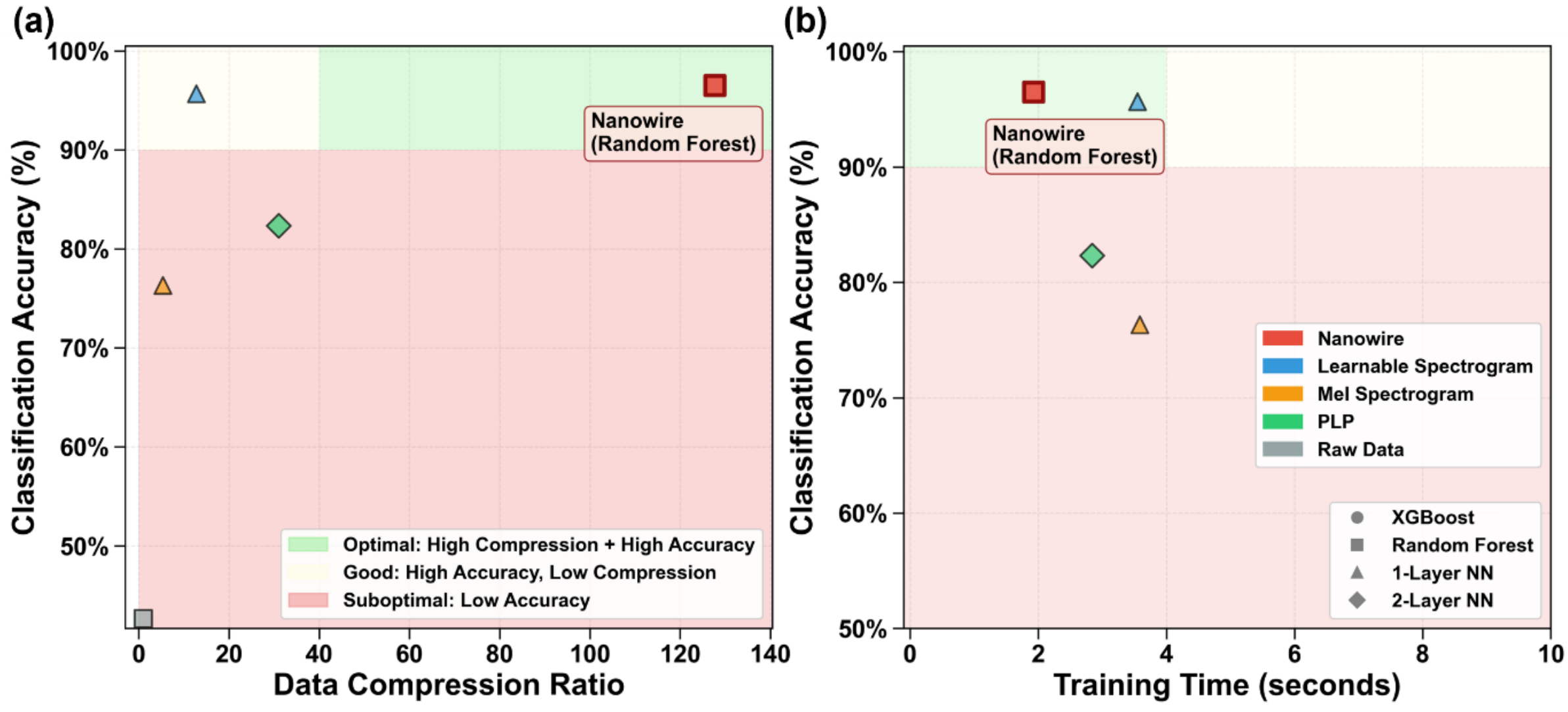

*Figure 5. Combined analysis of classification accuracy, data compression ratio, and training efficiency across feature representations and classifiers for multispeaker dataset. (a) Nanowire-processed features outperform all other software-preprocessing features, attaining 96.5% accuracy while achieving far superior data compression ratio. (b) Classification Accuracy vs Training Time comparison: Nanowire-features achieve higher classification accuracy while having the lowest training time (1.92s) due to their small feature set size.*

## 4. Discussion

This study demonstrates that memristive nanowire networks (NWNs) function as a highly efficient neuromorphic preprocessing layer for spoken-digit classification. NWNs achieve near state-of-the-art accuracy while compressing audio features by up to ~255 times, requiring as few as 16 input dimensions compared to ≥132 features used by conventional software techniques (Figure 4, Table 1). These compact representations enable sub-second training across multiple classifier architectures and maintain high accuracy in both single- and multi-speaker datasets, confirming classifier-agnostic performance and robust speaker generalization. Quantitatively, nanowire preprocessing achieves 96% feature reduction, much more than the best software alternative in multispeaker tasks (32 vs. 768 dimensions), while slightly improving accuracy (~0.9%) and reducing training time by ~46%. In single-speaker tasks, nanowires nearly reach the peak state-of-the-art accuracy (~99%) while delivering >4 times higher compression, highlighting their ability to produce compact yet highly discriminative representations.

Two intrinsic properties of NWNs underlie these advantages. First, the network inherently suppresses noise in raw audio signals: potentiation and depression rates at memristive junctions modulate the excitatory response, producing a tuneable low-pass filtering effect. This removes high-frequency components that obscure class-relevant patterns while retaining dominant signal structures, functioning analogously to a hardware moving average filter.

Second, the NWN implements a high-dimensional spatio-temporal transformation. Temporal inputs are projected into a 1024-dimensional conductance space, where recurrent connections integrate short- and long-term dynamics. Even single-terminal readouts encode cumulative information over time, enhancing linear separability and discriminative power. This mechanism robust performance in non-linear classifiers, with XGBoost and Random Forest achieving 98.95% and 97.9% accuracy at ~66 times and 255 times compression, respectively. Further investigation showed that improvements in accuracy are also observed when using nanowire data with linear classifiers (SI3, SI4) for both single and multispeaker dataset Uniform subsampling further distills these high-dimensional representations into compact yet informative feature sets, optimizing the trade-off between compression, accuracy, and training efficiency.

Crucially, NWNs also enhance speaker generalization. On multi-speaker datasets, nanowire-processed features consistently outperform raw and conventional software features, achieving up to 96.5% accuracy with the lowest training time (1.92 s) and highest compression (125 times). This demonstrates the network's ability to retain discriminative features across previously unseen speakers, making it particularly suitable for real-world, edge-deployed voice recognition systems that must operate under strict latency and resource constraints.

Collectively, these results establish NWNs as a powerful hybrid neuromorphic preprocessing approach. By combining noise suppression, high-dimensional spatio-temporal projection, and targeted feature subsampling, NWNs deliver compact, speaker-agnostic representations that enable rapid and accurate classification. Future work could explore multi-terminal readouts and adaptive potentiation strategies to further exploit network dynamics, potentially exceeding current state-of-the-art performance while preserving energy efficiency and low latency.

## 5. Conclusion

In this paper, we devised a neuromorphic computing approach using a nanowire network and software classification models for audio classification tasks. Prior works in this direction have explored neuromorphic computing approaches for associative learning and practical tasks such as MNIST image classification using reservoir computing, resistive switching memory, non-volatile random-access memory etc. Audio signal classification has been studied with spiking neural networks or with Mel Spectogram as a pre-processing technique. However, such pre-processing techniques are power hungry and bottleneck the latency benefits of using neuromorphic computing approaches. This is one of the first studies that leverages the spatio-temporal transformation properties of nanowire networks on audio signal classification without performing software pre-processing of data.

This study establishes memristive nanowire networks as compact, energy-efficient neuromorphic preprocessors that combine near–state-of-the-art classification accuracy with data compression far exceeding conventional software methods. Unlike digital preprocessing techniques, nanowire networks perform noise suppression and spatio-temporal transformation directly in hardware, extracting features that are both compact and highly informative. These features achieve 98.95% classification accuracy with ~66 times compression and 97.9% accuracy with an unprecedented ~255 times compression and sub-second training, outperforming Mel Spectrogram, PLP, and Learnable Spectrogram in data compression efficiency. The improvement in linear separability and consistent multi-speaker performance confirm that the extracted features are robust and generalizable. These results highlight nanowire networks as a scalable hardware foundation for low-power, low-latency machine learning pipelines, moving neuromorphic computing toward practical deployment.

The present study only considered the short-term memristive effects of the nanowire network since each audio file was processed through a new network. Future studies can investigate the long-term memristive effects of the network over multiple audio files, allowing the network to exhibit prediction capabilities, further exploiting the learning capabilities of the network. Memristive NWN is a rich system which offers multiple tuning parameters for effective neuromorphic computing such as time scales of synapses, number of connections, strength of connections and plasticity mechanism (directed by the materials used to realize the network). Future studies can also explore the optimization of multiple other parameters of the nanowire network such as wire density and using multiple output electrode measurements to study potential classification improvements and computation time reductions. The future scope of this research also includes further studies to validate the generalizability of nanowire extracted features among different genders, races and age-groups. Noisy signals can be better analysed by this approach; this approach can be applied when data storing and sending is an issue for edge computing use cases.

This theoretical modelling and simulation-based study of memristive nanowire networks provide guidelines for the fabrication of devices which opens avenues for the real time processing of audio signal. The present work is a proof of concept for realizing memristive NWN based edge computing devices.

**Acknowledgments**

All the authors are grateful for the project grant from the Ministry of Education (MOE), Government of Singapore "MOE-2019-T2-1-197" towards the funding of this project. Ankush Kumar acknowledges

Department of Science and Technology, India for the funding.

**Conflict of Interest**

There are no conflicts to declare.

**CRediT Statement**

Akshaya Rajesh: Conceptualization, Methodology, Software, Validation, Formal Analysis, Investigation, Data Curation, Writing-Original Draft, Writing-Review & Editing, Visualization. Pavithra Ananthasubramanian: Conceptualization, Methodology, Investigation, Data Curation, Writing-Original Draft, Writing- Review & Editing, Visualization, Supervision. Nagarajan Raghavan: Resources, Writing-Review & Editing, Supervision, Project Administration, Funding Acquisition. Ankush Kumar: Conceptualization, Methodology, Software, Validation, Formal Analysis, Investigation, Resources, Writing-Review & Editing, Supervision.

**Data Availability**

The datasets generated during and/or analysed during the current study are available from the corresponding author on reasonable request.